# Fast Implementation of Morphological Filtering Using ARM NEON Extension

**Elena Limonova[1], Arseny Terekhin[2], Dmitry Nikolaev[3] and Vladimir Arlazarov[3]**
[1]*Moscow Institute of Physics and Technology, 141700, 9, Institutskiy per., Dolgoprudny, Russia.*
[2]*Institute for Information Transmission Problems, RAS, 127994, 19, Bolshoy Karetny per., Moscow, Russia.*
[3]*Institute for System Analysis, FRC CSC RAS, 117312, 9, pr. 60-letiya Oktyabrya, Moscow, Russia.*

**Abstract**

In this paper we consider speedup potential of morphological image filtering on ARM processors. Morphological operations are widely used in image analysis and recognition and their speedup in some cases can significantly reduce overall execution time of recognition. More specifically, we propose fast implementation of erosion and dilation using ARM SIMD extension NEON. These operations with the rectangular structuring element are separable. They were implemented using the advantages of separability as sequential horizontal and vertical passes. Each pass was implemented using van Herk/Gil-Werman algorithm for large windows and low-constant linear complexity algorithm for small windows. Final implementation was improved with SIMD and used a combination of these methods. We also considered fast transpose implementation of 8×8 and 16×16 matrices using ARM NEON to get additional computational gain for morphological operations. Experiments showed 3 times efficiency increase for final implementation of erosion and dilation compared to van Herk/Gil-Werman algorithm without SIMD, 5.7 times speedup for 8×8 matrix transpose and 12 times speedup for 16×16 matrix transpose compared to transpose without SIMD.

**Keywords:** SIMD, computational efficiency, matrix transpose, erosion, dilation.

## 1. INTRODUCTION

Computational efficiency is a very important aspect of computer vision technologies. Image processing and image recognition are rapidly developed and involve more and more complicated algorithms. At the same time computational power of devices for such programs has significantly changed. It becomes common to use computer vision technologies not only on desktop computers, but also on mobile devices and in embedded systems [1-3]. Due to this reason, we need to use all available resources to speedup considered algorithms to meet execution time requirements, that are important in industrial systems. The examples of such systems can be found in [4-6].

Modern central processing units (CPUs) provide a number of means of developing efficient software. Among them are multicore computing of independent tasks and SIMD (Single Instruction Multiple Data) extensions which allow parallel single-core data processing.

SIMD is computational architecture which allows vector or matrix operations. The instructions are processed sequentially, but each instruction operates on a vector of data, providing data parallelism [7].

Despite the fact that modern compilers support automatic vectorization of source code, they are not always able to do efficient optimization [8]. To deal with this issue, developers can use SIMD extensions directly during development. Intel x86 CPUs support SIMD via MMX, SSE and AVX extensions, ARM CPUs support VFP and NEON SIMD extensions.

The best candidates for SIMD vectorization are basic image processing functions, which deal with every pixel. Processing of each pixel is often mathematically simple and involve computations on groups of sequential pixels. Such kind of processing is easy to implement with the help of SIMD extension.

While there is a large number of such basic image processing functions, some of them are used in almost every image processing system, for example, filtration, rotation, transpose, downscale, etc. In this paper we consider transpose and such morphological filtering operations as erosion and dilation.

Image transpose is an important image processing operation. It is often used in filtration algorithms, because many of widely used filters are separable (for example, Gauss filter, box filter, different kinds of morphological filtering) [9]. Separability allows to express filter application as a composition of horizontal and vertical passes of one-dimensional filters. Vertical filter is applied to image columns and processes all pixels of image row in the same way. It can use SIMD efficiently in case pixels in a row are stored in neighboring locations of memory. Horizontal filter is applied to image rows and processes all pixels of image column in the same way. It is often not easy to speedup it with SIMD, so we can perform transpose of an image, SIMD-implemented vertical filter pass and one more transpose. It is important to have fast transpose implementation in this case, because it is performed many times. The example of matrix transpose for 4×4 matrices with 16-bit data can be found in ARM NEON documentation [10]. Intel x86 SSE2 provides macro _MM_TRANSPOSE4_PS for the same purposes. Moreover, there are papers considering matrix transpose with the help of Intel x86 AVX [11]. In this paper we show 8×8 and 16×16 matrix transpose implementation using NEON.

Erosion and dilation are morphological operations, which are widely used in image processing, because all morphological operations can be expressed as their composition. Thus, improving efficiency of erosion and dilation we can reduce processing time in a number of cases.

## 2. MORPHOLOGICAL OPERATIONS

Morphological operations are used to analyse and process geometrical structures of image. Each morphological operation takes two arguments: source image and structuring element. Structuring element is normally smaller than the source image. Let the size of structuring element be $w_x \times w_y$. It has defined anchor point which is normally in its center. When we compute erosion, structuring element is moved over the source image, and we compute the minimal pixel value in the area overlapped by structuring element. This minimum is the result of erosion at the pixel in the anchor point position of current structuring element application:

$$D(x,y) = min\{S(x-n+w_x/2, y-m+w_y/2), (n,m) \subseteq T\},$$

where $D(x, y)$ is destination image, $S(x, y)$ is source image, $T = [0, w_x - 1] \times [0, w_y - 1]$ - the region where structuring element is defined.

Dilation is computed similarly using maximum instead of minimum.

Other morphological operations, such as opening, closing, morphological gradient, can be expressed via erosion, dilation and arithmetical operations.

## 3. SIMD extension ARM NEON

One possible solution for computational efficiency issues is to use SIMD extensions. Even typical mobile CPUs with ARM architecture normally support SIMD extension NEON.

SIMD computational architecture is illustrated in the Fig. 1 [7]. Because automatic vectorization is often not effective developers can use SIMD instructions directly via compiler intrinsics. Intrinsics are compiler primitives, that are handled in a special way. Because a compiler knows operations performed by intrinsics, it can optimize it efficiently and generate better instruction sequence than for normal function call. Moreover, intrinsics often can be more efficient than inline assembly, because the compiler can manage registers and execute some instructions out of order.

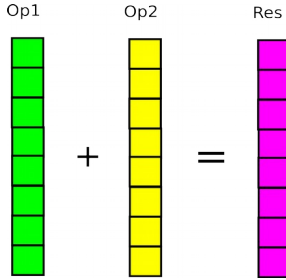

**Figure 1.** Addition of two vectors with 8 elements each using one SIMD addition.

Register size in ARM NEON is 128 bits. One can put 4 32-bit values, 8 16-bit values or 16 8-bit values in such register. ARM NEON includes a number of operation on 128-bit registers, for example, loading/storing data in memory, arithmetical operations, type conversion, element permutation, etc [12].

## 4. MATRIX TRANSPOSE WITH THE HELP OF INTRINSICS

Efficient matrix transpose implementation should consist of the minimal number of instructions. ARM instruction set provides special assembly instructions VTRN.n suitable for matrix transpose. VTRN.n instruction interprets operands as vectors with elements of n bits and perform transpose for 2×2 matrices constricted from these elements. The Fig. 2 shows VTRN.16 (n=16) instruction as an example.

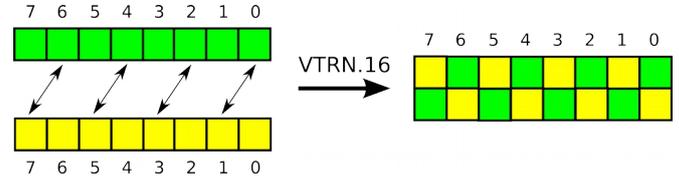

**Figure 2.** The result of VTRN.16 instruction.

Assembly instruction VTRN.n can be used in a flexible way via ARM NEON `vtrnq` intrinsics, that are converted to VTRN instructions by a compiler.

Let us define N×M.k matrix as the matrix of size N×M with elements of k bits each. ARM NEON documentation demonstrated 4×4.16 matrix transpose with 3 `vtrnq` intrinsics [10]. Similarly 4×4.32 matrix can be transposed using 4 `vtrnq` intrinsics: firstly, blocks of 2×2.32 are transposed in two first lines, then blocks of 2×2.32 are transposed in third and fourth lines and finally blocks of 2×2.64 are transposed.

We implemented also matrix transpose for 8×8.16 and 16×16.8 matrices. Transpose of 8×8.16 matrix was performed in 64 instructions: 16 load/store instruction, 32 data permutation instructions and 16 auxiliary instructions to interpret a vector of one type as a vector of other type. These auxiliary instructions are used for correct compilation and do not affect efficiency. The final transpose algorithm:

```
uint16x8x2_t t0 = vtrnq_u16(vld1q_u16(addr_s0), vld1q_u16(addr_s1));
uint16x8x2_t t1 = vtrnq_u16(vld1q_u16(addr_s2), vld1q_u16(addr_s3));
uint16x8x2_t t2 = vtrnq_u16(vld1q_u16(addr_s4), vld1q_u16(addr_s5));
uint16x8x2_t t3 = vtrnq_u16(vld1q_u16(addr_s6), vld1q_u16(addr_s7));

uint32x4x2_t x0 = vtrnq_u32(vreinterpretq_u32_u16(t0.val[0]),
vreinterpretq_u32_u16(t1.val[0]));
uint32x4x2_t x1 = vtrnq_u32(vreinterpretq_u32_u16(t2.val[0]),
vreinterpretq_u32_u16(t3.val[0]));
uint32x4x2_t x2 = vtrnq_u32(vreinterpretq_u32_u16(t0.val[1]),
vreinterpretq_u32_u16(t1.val[1]));
uint32x4x2_t x3 = vtrnq_u32(vreinterpretq_u32_u16(t2.val[1]),
vreinterpretq_u32_u16(t3.val[1]));

vst1q_u16(addr_d0, vreinterpretq_u16_u32(
vcombine_u32(vget_low_u32(x0.val[0]), vget_low_u32(x1.val[0]))));
vst1q_u16(addr_d1, vreinterpretq_u16_u32(
vcombine_u32(vget_low_u32(x2.val[0]), vget_low_u32(x3.val[0]))));
vst1q_u16(addr_d2, vreinterpretq_u16_u32(
vcombine_u32(vget_low_u32(x0.val[1]), vget_low_u32(x1.val[1]))));
```

```
vst1q_u16(addr_d3, vreinterpretq_u16_u32(
vcombine_u32(vget_low_u32(x2.val[1]), vget_low_u32(x3.val[1]))));
vst1q_u16(addr_d4, vreinterpretq_u16_u32(
vcombine_u32(vget_high_u32(x0.val[0]), vget_high_u32(x1.val[0]))));
vst1q_u16(addr_d5, vreinterpretq_u16_u32(
vcombine_u32(vget_high_u32(x2.val[0]), vget_high_u32(x3.val[0]))));
vst1q_u16(addr_d6, vreinterpretq_u16_u32(
vcombine_u32(vget_high_u32(x0.val[1]), vget_high_u32(x1.val[1]))));
vst1q_u16(addr_d7, vreinterpretq_u16_u32(
vcombine_u32(vget_high_u32(x2.val[1]), vget_high_u32(x3.val[1]))));
```

This implementation is 5.7 times faster than implementation without SIMD. The results of time measurements on ARM CPU Samsung Exynos 5422 with NEON support are shown in Table 1.

Transpose of 16×16.8 matrix was performed in the same way in 152 instructions (32 load/store instruction, 72 data permutation instructions and 48 auxiliary instructions to interpret a vector of one type as a vector of other type). This implementation is 12 times faster than implementation without SIMD (see Table 1).

**Table 1.** Execution time of matrix transpose on Samsung Exynos 5422 with ARM NEON.

| Matrix size | Data type | Execution time, without SIMD, ns | Execution time, with SIMD, ns |
| --- | --- | --- | --- |
| 8 × 8 | 16-bit unsigned int | 114 | 20 |
| 16 × 16 | 8-bit unsigned int | 565 | 47 |

## 5. FAST MORPHOLOGY IMPLEMENTATION

Erosion and dilation allow separable implementation, which has lower computational complexity, than non-separable one. We will consider such implementation with sequential vertical (with structural element of size $w_x \times 1$) and horizontal (with structural element of size $1 \times w_y$) passes. Let us consider two implementations of each pass. In both cases we suppose that image type is 8-bit unsigned integer.

### 5.1. Implementation of horizontal pass
#### 5.1.1. Baseline implementation

In this implementation of horizontal pass of erosion (dilation) we used van Herk/Gil-Werman algorithm to find minimum (maximum) on ranges of constant length $w_y$. It uses additional memory, which size is equal to doubled image size. This way we can solve the problem for each column, and the complexity will be linear in image size. Pixels in each row are processed in the same way, which allows us to use intrinsic `vminq_u8` (`vmaxq_u8`) to find minimum (maximum) of 16 pairs of 8-bit values in one instruction.

#### 5.1.2. Linear implementation

This implementation of horizontal pass of erosion (dilation) has linear complexity in window size $w_y$ and uses intrinsics. The source code in C++ language is below.

```
//uint8_t **src_lines - two-dimensional array of size width x height with 8-bit
unsigned int data, representing source image
//uint8_t **dst_lines - two-dimensional array of size width x height with 8-bit
unsigned int data, representing resulting image
//int wing - "wing" of structural element size, wy = 2*wing+1

for (int y = wing; y < height-wing-1; y += 2)
{
  for (int x = 0; x < width; x += 16)
  {
    uint8x16_t val = vld1q_u8(src_lines[y-wing+1] + x);
    for (int k = -wing + 2; k <= wing; k++)
      val = vminq_u8(val, vld1q_u8(src_lines[y+k]+x));
    vst1q_u8(dst_lines[y]+x, vminq_u8(val, vld1q_u8(src_lines[y-wing]+x)));
    vst1q_u8(dst_lines[y+1]+x,vminq_u8(val, vld1q_u8(src_lines[y+wing+1]+x)));
  }
}
```

Here we fill two sequential rows of resulting image by computing minimum (maximum) in a `for` loop. It can be noted, that for each pair of vertically adjacent rows there are $w_y$-2 elements that are common for range minimum (maximum) computing. This fact allowed us to optimize computations. In addition, we used intrinsics, because different elements in image rows are processed independently. Image edges were processed separately.

#### 5.1.3. Time measurements

Our experiments were performed on Samsung Exynos 5422 CPU with ARM architecture. We used gcc compiler and gcc intrinsics. Input was gray image of width of 800 pixels and height of 600 pixels with 8-bit unsigned integer data. Structuring element size was $1 \times w_y$, $w_y$ parameter varied. Experiment results are shown in the Fig. 3. Erosion and dilation execution times are identical, so we show only erosion time.

Usage of SIMD allowed us to speed up van Herk/Gil-Werman implementation more than 3 times. Linear implementation for $w_y=3$ is 14 times faster than van Herk/Gil-Werman without SIMD, however with $w_y$ raising the speedup decreases. We can see, that for $w_y \leq w_y^0$, $w_y^0 = 69$ linear implementation works efficiently, in case of bigger $w_y$ van Herk/Gil-Werman with SIMD performs better.

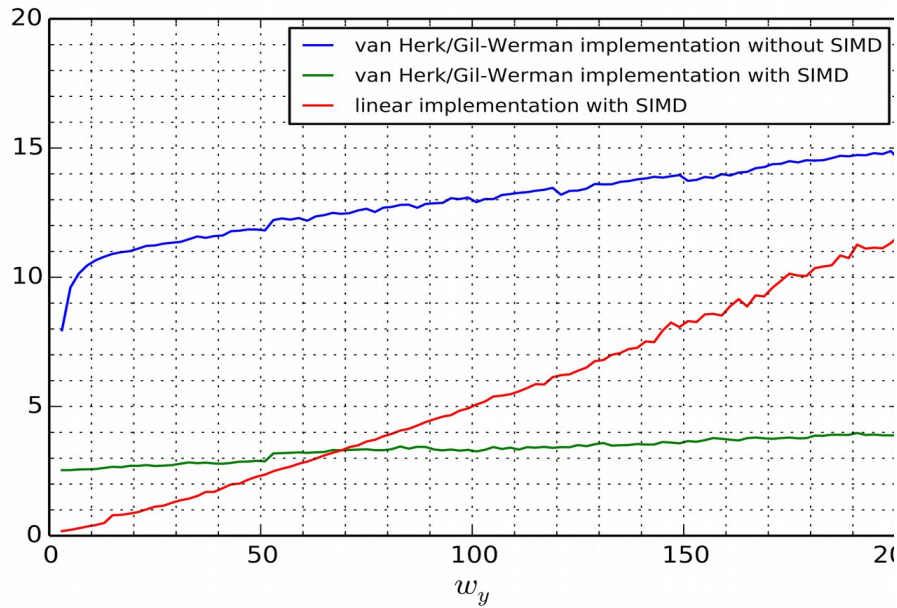

**Figure 3.** Execution time of different horizontal pass implementations of erosion with structuring element size of $1 \times w_y$.

**5.2. Vertical pass implementation**
**5.2.1. Baseline implementation**
We implemented vertical pass of erosion (dilation) using transpose and van Herk/Gil-Werman algorithm. Matrix transpose was implemented with the help of SIMD as in Section 4. Therefore, we use memory efficiently and take advantage of intrinsics.

**5.2. Linear implementation**
This implementation of vertical pass of erosion (dilation) is linear in window size $w_x$ and is improved with intrinsics. The source code in C++ language is below.

```
  //uint8_t **src_lines - two-dimensional array of size width x height with    8-bit
unsigned int data, representing source image
  //uint8_t **dst_lines - two-dimensional array of size width x height with 8-bit
unsigned int data, representing resulting image
  //int wing - "wing" of structural element size, wₓ = 2*wing+1
  for (int y = 0; y < height; ++y)
  {
    for (int x = 0; x < width; x += 16)
    {
      uint8x16_t val = vld1q_u8(src_lines[y]+x-wing);
      for (int j = x-wing+1; j <= x+wing; ++j)
        val = vminq_u8(val, vld1q_u8(src_lines[y]+j));
      vst1q_u8(dst_lines[y]+x, val);
    }
  }
```

Here we fill each row of resulting image by computing minimum (maximum) in a `for` loop of length $w_x$. With the help of intrinsics we can compute 16 window minimums (maximums) in $w_x$ instructions. Image edges were processed separately.

**5.2.3. Time measurements**
Our experiments were performed on Samsung Exynos 5422 with the gray image 800×600 pixels with 8-bit unsigned int data. We used gcc compiler and gcc intrinsics. Structuring element size was $w_x \times 1$, $w_x$ parameter varied. Experiment results are shown in the Fig. 4.

Usage of SIMD allowed us to speed up van Herk/Gil-Werman implementation almost 3 times for $w_x \geq 3$. Linear implementation for $w_x=3$ is 11 times faster than van Herk/Gil-Werman without SIMD, however with $w_x$ raising the speedup decreases. We can see, that for $w_x \leq w_x^0$, $w_x^0 = 59$ linear implementation works efficiently, in case of bigger $w_x$ van Herk/Gil-Werman with SIMD performs better.

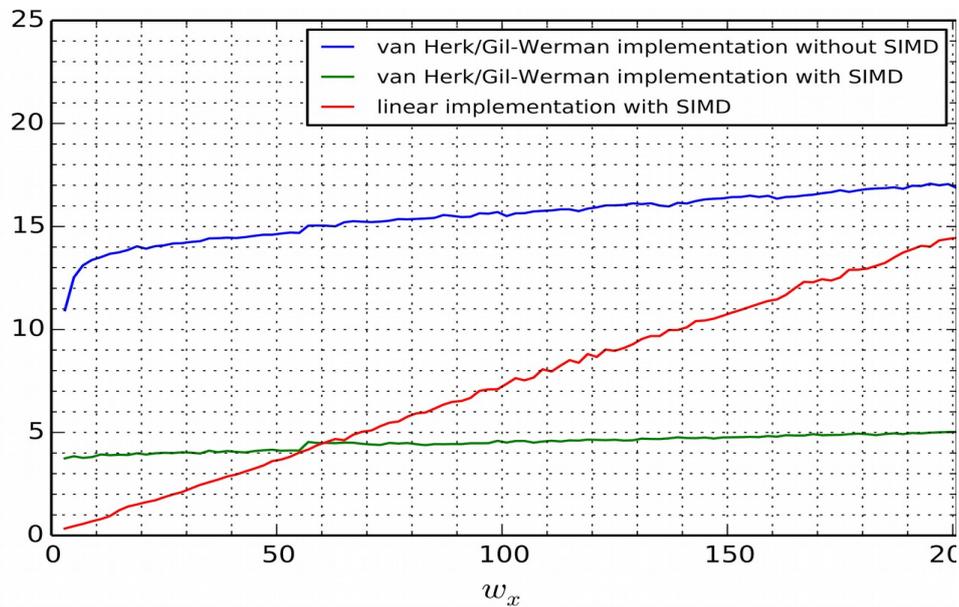

**Figure 4.** Execution time of different vertical pass implementations of erosion with structuring element size of $w_x \times 1$.

### 5.3. Final fast morphology implementation

Based on our experiments, the most efficient implementation of erosion (dilation) is a combination of considered methods. For the horizontal pass linear implementation is used, when $w_y \leq w_y^0$, $w_y^0 = 69$ and van Herk/Gil-Werman algorithm implementation with SIMD otherwise. For the vertical pass linear implementation is used, when $w_x \leq w_x^0$, $w_x^0 = 59$ and van Herk/Gil-Werman implementation with SIMD otherwise. Values $w_x^0$ and $w_y^0$ are different, because passes work with memory asymmetrically.

### CONCLUSION

Our experiments on processor with ARM NEON showed that SIMD can significantly improve morphological image filtering efficiency. With the help of SIMD we managed to speedup morphological erosion and dilation. We used separable implementation and improved efficiency of horizontal and vertical passes with gcc intrinsics. We considered two implementations for both passes. Baseline one used van Herk/Gil-Werman algorithm for both passes with two additional transposes for vertical pass. Alternative one had linear complexity in structuring element size. Final implementation was a combination of two considered methods. We obtained 3 times speedup relative to implementation of van Herk/Gil-Werman algorithm without SIMD. Thus, modern CPUs of ARM architecture have good opportunities for computational efficiency optimization of image processing algorithms.

### ACKNOWLEDGMENTS
Applied scientific research is supported by Ministry of Education and Science of Russian Federation (project RFMEFI58114X0003).

### REFERENCES

[1] A. Sheshkus, D. Nikolaev, A. Ingacheva, N. Skoryukina, "Approach to recognition of flexible form for credit card expiration date recognition as example", in: Proc. SPIE 9875, Eighth International Conference on Machine Vision, 98750L (December 8, 2015), 2015.

[2] E. Limonova, D. Ilin, D. Nikolaev, "Improving neural network performance on SIMD architectures", in: Proc. SPIE 9875, Eighth International Conference on Machine Vision, 98750L (December 8, 2015), 2015.

[3] F. Mamalet, S. Roux, C. Garcia, "Real-time video convolutional face finder on embedded platforms", EURASIP Journal on Embedded Systems, **1**, pp. 1-8, (2007).

[4] E. Kuznetsova, E. Shvets, D. Nikolaev, "Viola-Jones based hybrid framework for real-time object detection in multispectral images", in: Proc. SPIE 9875, Eighth International Conference on Machine Vision, 98750N (December 8, 2015), 2015.

[5] A. Mastov, I. Konovalenko, A. Grigoryev, "Application of random ferns for non-planar object detection", in: Proc. SPIE 9875, Eighth International Conference on Machine Vision, 98750M (December 8, 2015), 2015.

[6] V. Kopenkov, V. Myasnikov, "Detection and tracking of vehicles based on the video-registration information", in: Posters Proceedings of 23-rd International Conference on Computer Graphics, Visualization and Computer Vision (WSCG 2015), Czech Republic, Plzen, June 8 - 12, 2015, pp. 65–69.

[7] D. Patterson and J. Hennessy. Computer Organization and Design (4th Edition), Elsevier, (2009).

[8] J. Holewinski et. al., "Dynamic Trace-Based Analysis of Vectorization Potential of Applications," ACM SIGPLAN Notices 47(6), 371-382, (2012).

[9] R. Gonzalez, R. Woods. Digital Image Processing, 2nd Edition, Boston, USA, (2001).

[10] ARM Connected Community, Coding for NEON - Part 5: Rearranging Vectors, http://community.arm.com/groups/processors/blog/2012/03/13/coding-for-neon–part-5-rearranging-vectors.

[11] Ahmed S. Zekri. "Enhancing the matrix transpose operation using Intel AVX instruction set extension". International Journal of Computer Science & Information Technology (IJCSIT), **6**(3), pp.67-78, (2014).

[12] ARM Infocenter, ARM non-confidential Technical Publications, http://infocenter.arm.com/